# High-$Q$ coupled resonances on a PhC waveguide using a tapered nanofiber with high coupling efficiency

**Tomohiro Tetsumoto[1], Yuta Ooka[1], and Tanabe Takasumi[1,*]**

[1] *Department of Electronics and Electrical Engineering, Faculty of Science and Technology, Keio University,*
*3-14-1, Hiyoshi, Kohoku-ku, Yokohama 223-8522, Japan*
[*] *takasumi@elec.keio.ac.jp*

**Abstract:** We experimentally demonstrate high-$Q$ cavity formation at an arbitrary position on a silicon photonic crystal waveguide by bringing a tapered nanofiber into contact with the surface of the slab. An ultrahigh $Q$ of $5.1 \times 10^5$ is obtained with a coupling efficiency of 39%, whose resonant wavelength can be finely tuned by 27 pm by adjusting the contact length of the nanofiber. We also demonstrate an extremely high coupling efficiency of 99.6% with a loaded $Q$ of $6.1 \times 10^3$. In addition, we show that we can obtain an all-pass filter type coupled resonator system, which has the potential to be used for slow light generation.

## References and links

1. T. Tanabe, M. Notomi, E. Kuramochi, A. Shinya, and H. Taniyama, "Trapping and delaying photons for one nanosecond in an ultrasmall high-Q photonic-crystal nanocavity," Nat. Photonics **1**, 49-52 (2007).
2. B. Corcoran, C. Monat, and C. Grillet, D. J. Moss, B. J. Eggleton, T. P. White, L. O'Faolain, and T. F. Krauss, "Green light emission in silicon through slow-light enhanced third-harmonic generation in photonic-crystal waveguides," Nat. Photonics **3**, 206-210 (2009).
3. T. Baba, T. Kawasaki, H. Sasaki, J. Adachi, and D. Mori, "Large delay-bandwidth product and tuning of slow light pulse in photonic crystal coupled waveguide," Opt. Express **16**, 9245–53 (2008).
4. K. Nozaki, T. Tanabe, A. Shinya, S. Matsuo, T. Sato, H. Taniyama, and M. Notomi, "Sub-femtojoule all-optical switching using a photonic-crystal nanocavity," Nat. Photonics **4**, 477–483 (2010).
5. E. Kuramochi, K. Nozaki, A. Shinya, K. Takeda, T. Sato, S. Matsuo, H. Taniyama, H. Sumikura, and M. Notomi, "Large-scale integration of wavelength-addressable all-optical memories on a photonic crystal chip," Nat. Photonics **8**, 474–481 (2014).
6. T. Tanabe, H. Sumikura, H. Taniyama, A. Shinya, and M. Notomi, "All-silicon sub-Gb/s telecom detector with low dark current and high quantum efficiency on chip," Appl. Phys. Lett. **96**, 101103 (2010).
7. K. K. Mehta, J. S. Orcutt, O. Tehar-Zahav, Z. Sternberg, R. Bafrali, R. Meade, and R. J. Ram, "High-Q CMOS-integrated photonic crystal microcavity devices," Sci. Rep. **4**, 4077 (2014).
8. M. Shinkawa, N. Ishikura, Y. Hama, K. Suzuki, and T. Baba, "Nonlinear enhancement in photonic crystal slow light waveguides fabricated using CMOS-compatible process," Opt. Express **19**, 22208-18 (2011).
9. K. Srinivasan, P. Barclay, M. Borselli, and O. Painter, "Optical-fiber-based measurement of an ultrasmall volume high-Q photonic crystal microcavity," Phys. Rev. B **70**, 081306 (2004).
10. C. Grillet, C. Smith, D. Freeman, S. Madden, B. Luther-Davies, E. Magi, D. Moss, and B. Eggleton, "Efficient coupling to chalcogenide glass photonic crystal waveguides via silica optical fiber nanowires," Opt. Express **14**, 1070–8 (2006).
11. P. E. Barclay, K. Srinivasan, M. Borselli, and O. Painter, "Probing the dispersive and spatial properties of photonic crystal waveguides via highly efficient coupling from fiber tapers," Appl. Phys. Lett. **85**, 4 (2004).
12. C. Li, H. Zhang, M. Yu, and G. Q. Lo, "CMOS-compatible high efficiency double-etched apodized waveguide grating coupler," Opt. Express **21**, 7868–74 (2013).
13. V. R. Almeida, R. R. Panepucci, and M. Lipson, "Nanotaper for compact mode conversion," Opt. Lett. **28**, 1302–4 (2003).

14. K. Shiraishi, H. Yoda, A. Ohshima, H. Ikedo, and C. S. Tsai, "A silicon-based spot-size converter between single-mode fibers and Si-wire waveguides using cascaded tapers," Appl. Phys. Lett. **91**, 141120 (2007).
15. M.-K. Kim, I.-K. Hwang, M.-K. Seo, and Y.-H. Lee, "Reconfigurable microfiber-coupled photonic crystal resonator," Opt. Express **15**, 17241–7 (2007).
16. M.-K. Kim, J.-Y. Kim, J.-H. Kang, B.-H. Ahn, and Y.-H. Lee, "On-demand photonic crystal resonators," Laser Photon. Rev. **5**, 479–495 (2011).
17. H.-J. Lim, C.-M. Lee, B.-H. Ahn, and Y.-H. Lee, "Dual-rail nanobeam microfiber-coupled resonator," Opt. Express **21**, 6724-32 (2013).
18. MIT Photonic Bands (MPB) is a free software package for computing the band structures (dispersion relations) and electromagnetic modes of periodic dielectric structure. MPB was developed at MIT, http://abi-nitio.mit.edu/wiki/index.php/MIT_Photonic_Bands.
19. Y. Takahashi, H. Hagino, Y. Tanaka, B. Song, T. Asano, and S. Noda, "High- Q nanocavity with a 2-ns photon lifetime," Opt. Express **15**, 17206–13 (2007).
20. MEEP is a free finite-difference time-domain (FDTD) simulation software package developed at MIT, http://abi-nitio.mit.edu/wiki/index.php/Meep
21. C. P. Michael, M. Borselli, T. J. Johnson, C. Chrystal, and O. Painter, "An optical fiber-taper probe for wafer-scale microphotonic device characterization," Opt. Express **15**, 4745–52 (2007).
22. H. Sumikura, E. Kuramochi, H. Taniyama, and M. Notomi, "Ultrafast spontaneous emission of copper-doped silicon enhanced by an optical nanocavity," Sci. Rep. **4**, 5040 (2014).
23. V. Savona, "Electromagnetic modes of a disordered photonic crystal," Phys. Rev. B **83**, 079907 (2012).
24. S. Mazoyer, J. P. Hugonin, and P. Lalanne, "Disorder-induced multiple scattering in photonic-crystal waveguides," Phys. Rev. Lett. **103**, 063903 (2009).
25. M. Patterson and S. Hughes, "Interplay between disorder-induced scattering and local field effects in photonic crystal waveguides," Phys. Rev. B **81**, 245321 (2010).
26. H. Thyrrestrup, S. Smolka, L. Sapienza, and P. Lodahl, "Statistical theory of a quantum emitter strongly coupled to Anderson-localized modes," Phys. Rev. Lett. **108**, 113901 (2012).
27. M. Notomi, E. Kuramochi, and T. Tanabe, "Large-scale arrays of ultrahigh-Q coupled nanocavities," Nat. Photonics **2**, 741–747 (2008).
28. A. Melloni, A. Canciamilla, C. Ferrari, F. Morichetti, L. O'Faolain, T. F. Krauss, R. D. La Rue, A. Samarelli, and M. Sorel, "Tunable delay lines in silicon photonics: Coupled resonators and photonic crystals, a comparison," IEEE Photonics J. **2**, 181-194 (2010).
29. F. Xia, L. Sekaric, and Y. Vlasov, "Ultracompact optical buffers on a silicon chip," Nat. Photonics **1**, 65–71 (2007).

## 1. Introduction

Photonic crystal (PhC) devices [1] are prominent candidates for optical signal processing because of their suitability for integration and their potential for strong light-matter interaction. PhCs have been used to realize many devices including slow light waveguides with a large group velocity [2, 3], an optical switch for ultra-low energy consumption [4], a wavelength-addressable all-optical memory [5], and a high quantum efficiency photodetector [6]. Furthermore, recent progress on fabrication technologies for CMOS compatible processes has enabled us to fabricate PhC devices using photolithography [7, 8], which will accelerate their industrialization. Although the coupling between a PhC nanocavity and waveguides can be well designed, the coupling efficiency between an optical fiber and a PhC/silicon waveguide is usually not high due to the mode mismatch between the waveguide and the optical fiber.

To overcome this problem, tapered nanofiber has been used to couple the light evanescently into the PhC devices [9]. A tapered fiber is fabricated by heating and pulling an optical fiber. The narrowest point of the fiber has a sub-micrometer radius; thus light can couple evanescently into the cavity. The coupling efficiency of tapered fiber is as high as 98% for a chalcogenide glass PhC waveguide [10] and 95% for a silicon PhC waveguide [11], and these values are even higher even if a grating coupler (maximum 70%) or a spot-size converter (maximum 90%) is used at the silicon wire-waveguide end [12-14].

Recently, Y.-H. Lee *et al.* reported the demonstration of a reconfigurable fiber-coupled PhC nanocavity [15, 16]. In addition to the controllability of the coupling efficiency, this scheme also allowed control of the position and resonant wavelength of the cavity. However, the experimental quality factor ($Q$) of the cavity was limited to around $10^4$ when the cavity was formed on an InGaAsP/InP quantum dot PhC waveguide. Numerical calculations revealed that the coupling efficiency could be as high as 90%, but the demonstrated efficiency was only a few percent with a 2D PhC waveguide [16] and 30% with a 1D nanobeam waveguide [17].

Motivated by these highly relevant reports, in this study we formed a reconfigurable cavity on a two-dimensional silicon PhC waveguide with a tapered nanofiber for the simultaneous experimental demonstration of an ultrahigh $Q$ and high coupling efficiency. We will show that a $Q$ of $5.1 \times 10^5$ is possible with a coupling efficiency of 39%. We also investigated the tuning performance of the resonant wavelength by changing the length of the contact area with the nanofiber. In addition, we demonstrated all-pass filter type multimode excitation, which may be used for the generation of slow light.

## 2. Theory of cavity formation

When a nanofiber approaches the top of a PhC waveguide, the effective refractive index of the waveguide increases; thus the cutoff frequency of the waveguide decreases and a modegap cavity is formed. We calculated the band diagram of the fiber-coupled silicon PhC waveguide by using MPB [18] with the following parameters; lattice constant $a = 420$ nm, waveguide width $w = 0.98\sqrt{3}a$ (W0.98), hole radius $r = 0.30a$, slab thickness $t = 0.50a$ and refractive index of silicon $n_{Si} = 3.47$. The radius of the silica fiber was 500 nm ($n_{SiO2} = 1.45$). The result is shown in Fig. 1(a). When we reduced the gap between the fiber and the waveguide, the cutoff frequency of the PhC waveguide moved downward. This localized downshift of the cutoff frequency allowed modegap confinement and made it possible to form a cavity [1, 19].

To confirm the cavity formation, we calculated the mode profile of the fiber-coupled PhC nanocavity with a 3-D finite difference time domain (FDTD) method [20]. We modeled a bent nanofiber with a curvature radius of $R = 125$ μm. The cross-sectional radius of the nanofiber was $r = 500$ nm. When we brought the fiber into contact with a W0.98 PhC waveguide we obtained a mode profile as shown in Fig. 1(b). The $Q$ value was $1.4 \times 10^7$ and the mode volume was $1.9(\lambda/n)^3$.

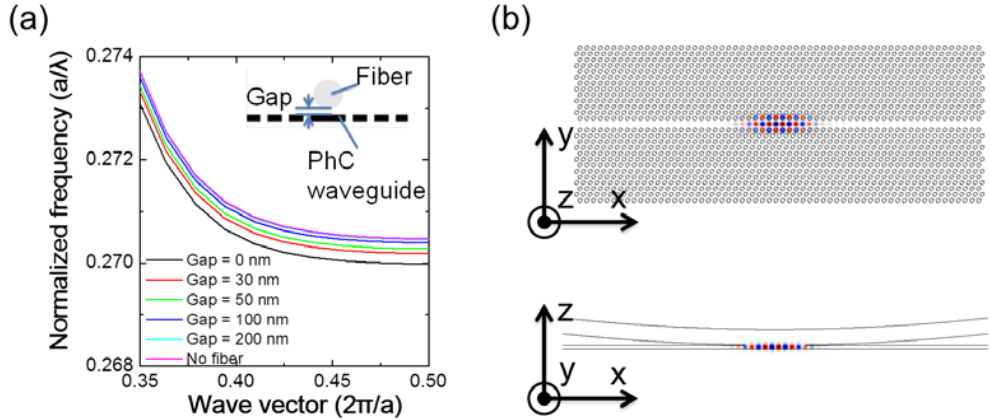

Fig. 1. (a) Band diagram of PhC waveguides in contact with a nanofiber. (b) Calculated $H_z$ field profile of an optical cavity created with a nanofiber. The upper and lower figures are views from the top and side, respectively.

## 3. Experiment

*3.1 Demonstration of ultrahigh-Q cavity formation*

In this section, we describe our experimental demonstration of cavity formation on a PhC waveguide. For the experiment, we used a dimpled tapered nanofiber [21] that we fabricated with the following processes. First, we fabricated a tapered nanofiber by heating and pulling a standard single-mode optical fiber. Next, we prepared a bare optical fiber with a radius of 62.5 μm as a mold and brought it into contact with the tapered fiber. Then, we heated the contact area to form a dimple in the nanofiber. The fabricated dimpled fiber is shown in Fig. 2(a). This bent and tensed fiber is useful for coupling light into a PhC waveguide from the top of the slab and enables us to control of the size of the contact area precisely. The insertion loss of our fiber was 10 dB; however, it can be reduced to almost 0 dB according to a previous report [21].

To prepare the PhC waveguide, we used a silicon foundry service to fabricate a silica-clad PhC waveguide with a 248-nm photolithography process. Then the silica cladding was removed by wet etching with 20% hydrogen fluoride.

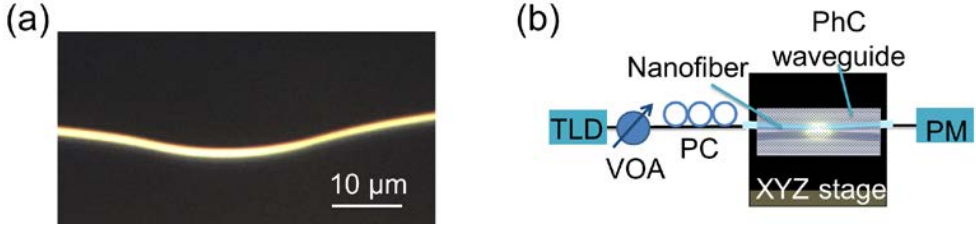

Fig. 2. (a) Fabricated dimpled fiber. The diameter is about 500 nm and the transmittance is −10 dB. (b) Experimental setup. TLD: Tunable laser diode. VOA: Variable optical attenuator. PC: Polarization controller. PM: Power monitor.

The optical measurement was performed by bringing the dimpled fiber into contact with the PhC waveguide and recording the transmittance spectrum with the setup shown in Fig. 2(b). The measured transmittance spectrum is shown in Fig. 3(a), where the cavity resonances are observed as dips, because we used a side-coupled configuration. The reason for the multiple resonances will be discussed later. When we measured the spectral width of one of the dips, we observed a loaded $Q_l$ of $5.1 \times 10^5$ as shown in the inset of Fig. 3(a). This resonance had a transmittance $T$ of 61% (i.e. a dip depth of 39%, which corresponds to the coupling efficiency), with which we obtained an unloaded $Q_u$ of $6.4 \times 10^5$.

To confirm the localization of the mode, we increased the input power and observed the influence of the thermo-optic (TO) effect, which is caused by multiphoton absorption if the light is strongly localized. As the result in Fig. 3(b) shows, clear TO bistability was observed, which is strong evidence of light localization. Another direct proof of the localization was the observation of the localized mode from the top of the slab achieved by using an infrared image sensor. Figure 3(c) contains images showing an input laser light when the cavity is on resonance and off resonance. We clearly observed a localized light in the contact region of the nano-tapered dimpled fiber.

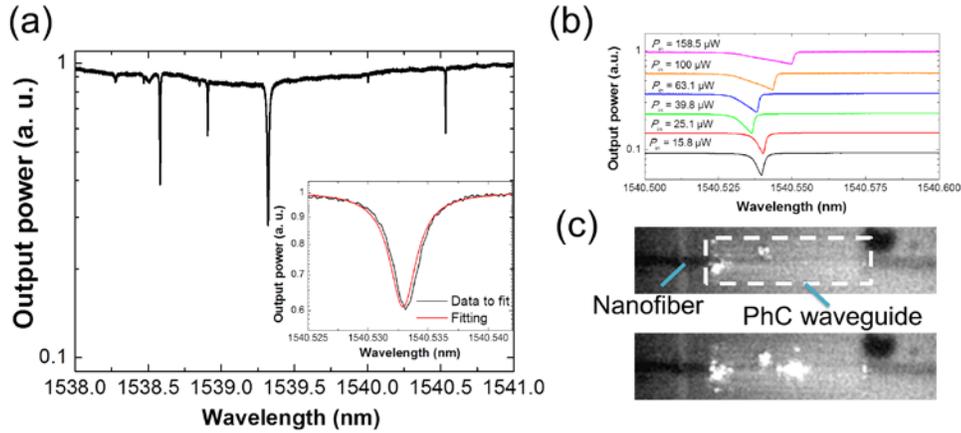

Fig. 3. (a) Transmittance spectrum of a reconfigurable fiber coupled PhC cavity.
(b) Transmittance spectra of the fiber coupled PhC cavity at different input powers $P_{in}$. $P_{in}$ is
the power in the nano-tapered fiber immediately before the contact point.   (c) Infrared image
of a reconfigurable fiber coupled PhC cavity. The upper image is when the cavity is
off-resonance. The lower is an image of a cavity on-resonance (Input wavelength of
1538.90 nm). The bright spot at the center of the image is the localized mode.

Next, we describe the polarization dependence. Since only the transverse electric field (TE) mode exhibits a photonic band-gap, the observation of the polarization dependence is the third proof of the excitation of the mode-gap resonances. We show the transmittance spectra at two different polarizations (90 deg) in Fig. 4(a), where we observed resonances only when we excited the cavity in the TE mode.

Finally, we attempted to maximize the coupling efficiency by carefully controlling the polarization. A maximized coupling efficiency of 96.4% was obtained for a mode with a $Q_l$ of $4.9 \times 10^4$ at wavelength of 1538.29 nm, and a coupling efficiency of 99.6% was achieved for a mode with $Q_l$ of $6.1 \times 10^3$ at a wavelength of 1535.37 nm (Fig. 4(b)).   Such a high coupling efficiency directly into an optical fiber is attractive for various applications including switching and quantum optics.

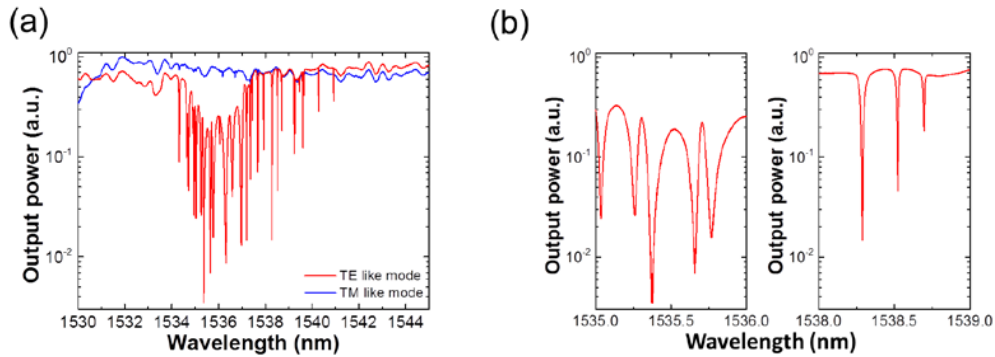

Fig. 4. (a) Transmittance spectrum of nearly orthogonal polarization light. (b) Enlarged views
of a mode with maximized coupling efficiency.

*3.2 Tuning the resonant wavelength*

Next, we demonstrate control of the resonant wavelength of the cavity by changing the length of the contact area between a nanofiber and a PhC waveguide. We moved the xyz translation stage and shifted the PhC waveguide 100 nm downwards from the initial contact position (Fig. 5(a)). As the stage moves downwards, the contact area between the nanofiber and PhC waveguide gradually decreases. This changes the size of the contact area and results in a blue shift of the resonant wavelength because of the shorter cavity length. The experimental result is shown in Fig. 5(b). The resonant wavelength shift is almost uniform for different modes. The wavelength tuning resolution is about 27 pm.

The high $Q$ and the high/direct coupling characteristics of this cavity, along with the wavelength tuning ability and the re-location ability, are extremely useful if we want to use such devices for strong and weak coupling experiments in, for example, cavity quantum electron-dynamics [22], in particular when we have a wavelength tuning precision of several tens of pm.

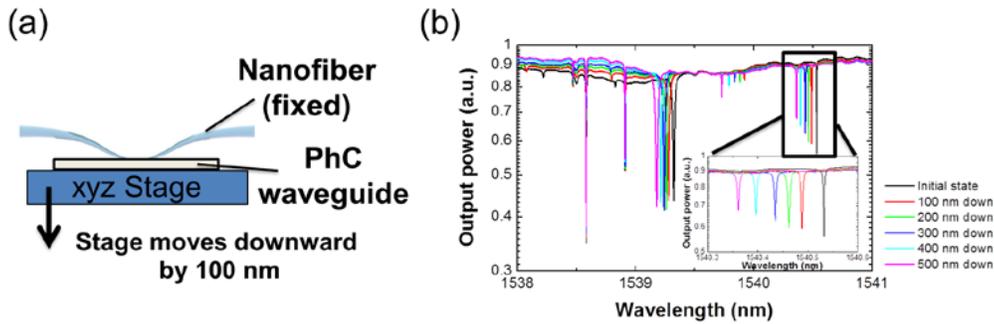

Fig. 5. (a) Schematic illustration of how we change the size of the contact area. (b) Spectral response of the resonant wavelength tuning using the method in (a).

**4. Demonstration of coupled cavity by utilizing waveguide disorder**

Finally, we focus on the TE-like mode and discuss the reason for the observed excitation of multi-resonant peaks. Figure 6(a) is the transmittance spectrum at a wavelength of around 1537 nm, where we observed at least 28 peaks in a 6.6 nm wavelength range. The wavelength separation between the peaks is too small to explain in terms of the excitation of different longitudinal modes of a Fabry-Pérot cavity.

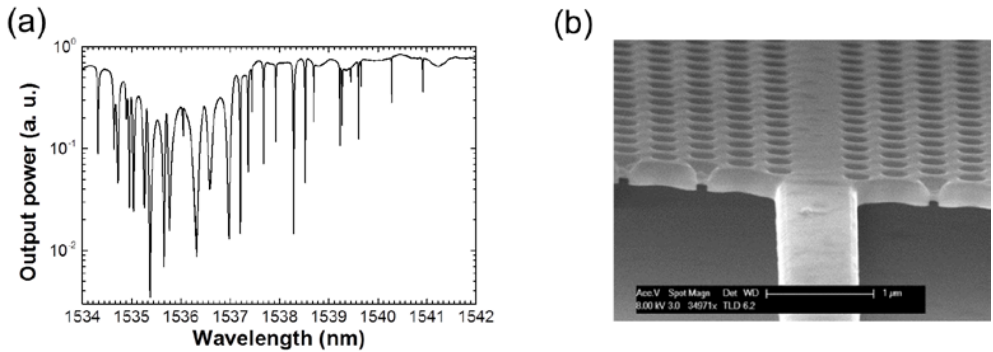

Fig. 6. (a) Transmittance spectrum of TE-like mode. (b) SEM image of PhC waveguide surface.

To clarify the origin of these peaks, we measured the surface of the PhC waveguide with a scanning electron microscope (SEM). The SEM image is shown in Fig. 6(b). We found that the surface of the PhC waveguide is not perfectly flat, but has bumps. These bumps result in a small fluctuation in the potential well along the waveguide, thus, multiple cavities are formed. The peaks at shorter wavelengths tend to have a lower $Q$ (the average $Q$ at shorter wavelengths is $2.8 \times 10^4$, in contrast to $1.4 \times 10^5$ at longer wavelengths), which we believe is due to the different spectral distances from the mode-gap of the continuum propagation mode of the waveguide. The multiple scattering that occurs in a disordered PhC waveguide has been well studied by a number of groups in connection with the Anderson localization of light, and the characteristics of the obtained spectrum are indeed close to those reported in these studies [23-26].

To investigate the above phenomenon in more detail, we monitored the mode profile of these localized modes with an infrared camera from top of the PhC waveguide slab as shown in Fig. 7. When we input light with a wavelength of 1534.33 nm, we observe multiple bright spots along the PhC waveguide, which is evidence of multiple resonances at different positions. Although a closer investigation of the coupling between the cavities is needed to clarify the characteristics as a coupled resonator optical waveguide (CROW) [27, 28], the measured image of the multiple bright spots suggests that the device may be used as an all-pass filter (APF) type coupled cavity system [29], which is useful for demonstrating optical buffers and slow light. It should be noted that these resonances can be excited with high coupling through a tapered fiber, which is the advantage of this scheme.

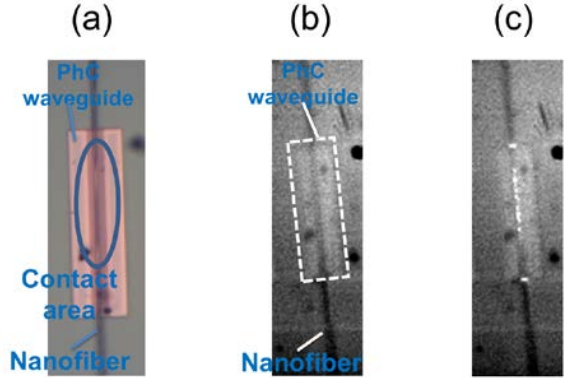

Fig. 7. Images of a fiber-coupled PhC cavity observed from the top of the slab. (a) Visible CCD image. (b) Infrared camera image at 1550 nm input (no resonance). (c) As (b) but at 1534.33 nm, which is in resonance with the structure.

## 4. Summary

In summary, we demonstrated experimentally high-$Q$ cavity formation on a Si PhC waveguide with a tapered nanofiber, and obtained an ultrahigh $Q$ of $5.1 \times 10^5$ at a coupling efficiency of 39%. These high values were obtained because we employed a silicon PhC waveguide as a platform. We also demonstrated high coupling efficiency and obtained an extremely high 96.4% transmittance for a mode with a $Q_l$ of $4.9 \times 10^4$ and an even higher 99.6% for a mode with a $Q_l$ of $6.1 \times 10^3$. We also achieved resonant wavelength control with a resolution of 27 pm.

In addition to our demonstration of a single cavity, we reported an APF type coupled cavity that originated from the surface disorder of a PhC waveguide. It may allow us to realize an optical buffer and slow light that have high coupling with optical fiber.

**Acknowledgment**

This work was supported by a Grant-in-Aid from the Ministry of Education, Culture, Sports, Science and Technology, Japan for the Photon Frontier Network Program.